\documentclass[twocolumn,prb,superscriptaddress,aps,showpacs]{revtex4}

\usepackage{graphicx}
\usepackage{dcolumn}
\usepackage{amsmath}
\usepackage{tikz}
\usepackage{subfigure}

\begin{document}
\title{Neutron Scattering Investigation of Rhenium Orbital Ordering in $3d-5d$ Double Perovskite Ca$_2$FeReO$_6$}
\date{\today}

\author{Bo Yuan}
\affiliation{Department of Physics, University of Toronto, Toronto, Ontario M5S~1A7, Canada}
\author{J. P. Clancy}
\affiliation{Department of Physics, University of Toronto, Toronto, Ontario M5S~1A7, Canada}
\author{J. A. Sears}
\affiliation{Department of Physics, University of Toronto, Toronto, Ontario M5S~1A7, Canada}
\author{A. I. Kolesnikov}
\affiliation{Neutron Scattering Sciences Division, Oak Ridge National Laboratory, Oak Ridge, Tennessee 37831, USA}
\author{M. B. Stone}
\affiliation{Neutron Scattering Sciences Division, Oak Ridge National Laboratory, Oak Ridge, Tennessee 37831, USA}
\author{Z. Yamani}
\affiliation{Chalk River Laboratories, National Research Council, Chalk River, Ontario K0J 1J0, Canada}
\author{Choongjae Won}
\affiliation{Department of Physics, Inha University, Incheon 402-751, Korea}
\author{Namjung Hur}
\affiliation{Department of Physics, Inha University, Incheon 402-751, Korea}
\author{B. C. Jeon}
\affiliation{Center for Correlated Electron Systems, Institute for Basic Science (IBS), Seoul 08826, Republic of Korea}
\affiliation{Department of Physics \& Astronomy, Seoul National University, Seoul 08826, Republic of Korea}
\author{T. W. Noh}
\affiliation{Center for Correlated Electron Systems, Institute for Basic Science (IBS), Seoul 08826, Republic of Korea}
\affiliation{Department of Physics \& Astronomy, Seoul National University, Seoul 08826, Republic of Korea}
\author{Arun Paramekanti}
\affiliation{Department of Physics, University of Toronto, Toronto, Ontario M5S~1A7, Canada}
\author{Young-June Kim}
\email{yjkim@physics.utoronto.ca}
\affiliation{Department of Physics, University of Toronto, Toronto, Ontario M5S~1A7, Canada}
\begin{abstract}

We have carried out inelastic neutron scattering experiments to study magnetic excitations in ordered double perovskite Ca$_2$FeReO$_6$. We found a well-defined magnon mode with a bandwidth of $\sim$50meV below the ferri-magnetic ordering temperature ($T_c\sim$520K), similar to previously studied Ba$_2$FeReO$_6$. The spin excitation is gapless for most temperatures within the magnetically ordered phase. However, a spin gap of $\sim$10meV opens up below $\sim$150K, which is well below the magnetic ordering temperature but coincides with a previously reported metal-insulator transition and onset of structural distortion. The observed temperature dependence of spin gap provides strong evidence for ordering of Re orbitals at $\sim$150~K, in accordance with earlier proposal put forward by Oikawa $\it{et.\,al}$ based on neutron diffraction [J. Phys. Soc. Jpn., $\bf{72}$, 1411 (2003)] as well as recent theoretical work by Lee and Marianetti [Phys. Rev. B, $\bf{97}$, 045102 (2018)]. The presence of separate orbital and magnetic ordering in Ca$_2$FeReO$_6$ suggests weak coupling between spin and orbital degrees of freedom and hints towards a sub-dominant role played by spin orbit coupling in describing its magnetism. In addition, we observed only one well-defined magnon band near magnetic zone boundary, which is incompatible with simple ferrimagnetic spin waves arising from Fe and Re local moments, but suggests a strong damping of Re magnon mode.

\end{abstract}
\maketitle
\section{Introduction}
Magnetism in $3d-5d$ ordered double perovskite (DP), A$_2$MM$^{\prime}$O$_6$, where A is a cation and M, M$^{\prime}$ are $3d$ and $5d$ transition metal ions, has been an area of intense research\cite{Ba2YOsO6, Ba2NaOsO6_ordered, Ba2MOsO6_ordered, Ba2CaMO6_order, IrDP_order,Sr2MgReO6_SG,Sr2CaReO6_SG,ReDP_singlet, Sr2BIrO6_SG_singlet,LCIO_order,Sr2CoOsO6_1,Sr2CoOsO6_2,Sr2FeOsO6, Sr2FeOsO6_1, Sr2CrOsO6, Sr2CrReO6, BFRO_OO, CFROstructuraldistortion}. Formed by alternately arranged $3d$ M and $5d$ M$^{\prime}$ ions in a (nearly) cubic lattice\citep{DP_review}, their magnetic properties may change dramatically depending on the identities of M and M$^{\prime}$ ions. When M-site is occupied by non-magnetic ions, $5d$ M$^{\prime}$ ions form a geometrically frustrated face-centered cubic lattice, which is shown to exhibit a variety of magnetic ground states ranging from magnetic and/or orbital order\cite{Ba2YOsO6, Ba2NaOsO6_ordered, Ba2MOsO6_ordered, Ba2CaMO6_order, IrDP_order, IrtheoryDP, d1theory, d2theory}, spin glass\cite{Sr2MgReO6_SG,Sr2CaReO6_SG} to exotic spin liquid phases.\cite{ReDP_singlet, Sr2BIrO6_SG_singlet} With magnetic M-site ions, one expects dominant super-exchange interaction between M and M$^{\prime}$ magnetic moments and a simple antiparallel arrangement between them. However, this simple picture is complicated by the presence of strong spin-orbit coupling (SOC) on $5d$ ions which tend to lock their spin and orbital angular momenta into the so-called $J_{eff}$ moments. Unlike isotropic Heisenberg interactions between spin only moments, interactions between $J_{eff}$ moments\cite{IrtheoryDP, d1theory, d2theory} and their interactions with spin only $3d$ moments\cite{Zn2FeOsO6_DM} both present in $3d-5d$ DP's can be anisotropic and strongly bond dependent. Competition between $5d-5d$ and $3d-5d$ interactions is behind the highly non-trivial magnetic order observed in a wide range of DP's\cite{LCIO_order,Sr2CoOsO6_1,Sr2CoOsO6_2,Sr2FeOsO6, Sr2FeOsO6_1, Sr2CrOsO6, Sr2CrReO6, BFRO_OO, CFROstructuraldistortion}.

An underlying assumption in the above discussion is the validity of $J_{eff}$ description of $5d$ ions. Since this is essentially an atomic description, where the physics is dominated by local energy scales such as SOC, crystal field and electronic correlation, one expects it to work well in strongly Mott-insulating DP's. On the other hand, one expects this picture to break down in metallic DP's where a large electronic bandwidth of $5d$ ions mixes different $J_{eff}$ levels and reduces the effects of SOC. It is interesting to ask whether the spin-orbit locked $J_{eff}$ picture arising from large SOC is still valid close to a Mott-instability, where both the correlated and itinerant nature of $5d$ electrons are important. Among $3d-5d$ DP's, A$_2$FeReO$_6$ (A=Ba,Sr,Ca with increasing lattice distortion) series provide examples that are likely to be close to a Mott instability. This is evidenced by a large change in transport properties across the series: cubic Ba$_2$FeReO$_6$ and tetragonal Sr$_2$FeReO$_6$ are metallic/half-metallic, while Ca$_2$FeReO$_6$ is insulating with a significant monoclinic lattice distortion.\citep{AFRO_transport} Among the three DP's in A$_2$FeReO$_6$ series, Ca$_2$FeReO$_6$ undergoes a thermally driven insulator to metal transition \citep{CFRO_MIT} at $\sim$150K, which hints towards its closest proximity to a Mott instability.

Theoretically, electronic properties of A$_2$FeReO$_6$ were recently investigated by DFT+U calculation by Lee and Marianetti\citep{CFRO_DFT3}. They showed that Hubbard term on Re sites in A$_2$FeReO$_6$ ($U_{Re}$) is very close to the critical value for a Mott transition. In contrast with the above local picture where SOC plays an important role, they found that lattice distortion and Re electron correlation were the determining factors for electronic properties of A$_2$FeReO$_6$ while SOC was less important. In particular, tilting of ReO$_6$ octahedra and large $U_{Re}$ enhance tendency for Re orbital order and lead to the insulating ground state of Ca$_2$FeReO$_6$. On the other hand, they predicted orbital order to be absent in less distorted Sr$_2$FeReO$_6$ which has a metallic ground state. Based on their calculation, such orbital order should also be absent in Ba$_2$FeReO$_6$, which is the least distorted member of A$_2$FeReO$_6$.

Structural evidence for orbital ordering in Ca$_2$FeReO$_6$ was reported in a neutron diffraction study by Oikawa $\it{et.\,al}$\cite{CFROstructuraldistortion}. They observed a slight change in distortion of ReO$_6$ and FeO$_6$ octahedra without any change in lattice symmetry across the metal-insulator transition. Specifically, they found that the octahedra went from a compressed one along $c$ at T$>$150K to an elongated one where the axis of elongation alternates between $a$ and $b$ for neighbouring Re sites at T$<$150K. In addition, they also observed a re-orientation of ordered moments from $c$-axis to $ab$-plane going from the high temperature to low temperature phase, which was argued to be a consequence of orbital ordering. Although structural distortion observed by Oikawa $\it{et.\,al}$\cite{CFROstructuraldistortion} is suggestive of orbital ordering, it alone cannot unambiguously confirm the presence of orbital order in Ca$_2$FeReO$_6$ as lattice distortion has also been observed in Ba$_2$FeReO$_6$ where the crystal goes from cubic to tetragonal below the magnetic ordering temperature\citep{BFRO_OO}, which according to Lee and Marianetti\cite{CFRO_DFT3} does not have orbital order.

Since magnetic interactions in Fe-Re double perovskites are strongly dependent on Re orbital states through SOC, another way to probe orbital ordering in these DP's is by measuring their magnetic excitations. Therefore, to examine orbital order in Ca$_2$FeReO$_6$ and gain a systematic understanding of magnetism in A$_2$FeReO$_6$, we studied magnetic excitations in Ca$_2$FeReO$_6$ using inelastic neutron scattering and compared it with previously studied Ba$_2$FeReO$_6$\cite{Plumb_BFRO}. We found that spin excitation of Ca$_2$FeReO$_6$ is quite similar to Ba$_2$FeReO$_6$ in that a gapless magnon mode dominates low energy region and the magnon bandwidth is $\sim$50meV. However, a large energy gap of $\sim$10meV opens up in the magnon spectrum for $T\lesssim$150K, which is well below the magnetic ordering temperature ($\sim$520K \cite{phase_separationCFRO}), but coincides with the proposed orbital ordering transition\cite{CFRO_MIT}. Opening of magnon gap directly indicates an increase of low symmetry magnetic interactions for $T\lesssim$150K, and therefore provides strong evidence for orbital ordering. On the other hand, gapless magnetic excitations were observed in Ba$_2$FeReO$_6$ that persist down to $\sim$30K indicating the absence of such an orbital ordering. We argue that separate magnetic and orbital ordering transitions in double perovskite Ca$_2$FeReO$_6$ as well as magnetic order without orbital order in Ba$_2$FeReO$_6$ suggest a separation between spin and orbital degrees of freedom in these double perovskites. This indicates a sub-dominant role of SOC in describing the magnetism of A$_2$FeReO$_6$. In addition, we observed only one magnon band near the magnetic zone boundary. We showed that this is inconsistent with ferrimagnetic spin-wave dispersions consisting of localized Fe and Re spins and instead suggests a coexistence of damped and undamped magnon modes.

\section{Experimental Details}
A polycrystalline sample of Ca$_2$FeReO$_6$ (8.9g) was synthesized using standard solid-state methods, as reported elsewhere.\cite{CFRO_DFT3, AFRO_transport} Magnetization measurements were carried out on pelletized polycrystalline sample using Quantum Design Magnetic Property Measurement System (MPMS).

Time-of-flight inelastic neutron measurements were performed using fine-resolution Fermi-chopper spectrometer (SEQUOIA) at the Spallation Neutron Source (SNS) at Oak Ridge National Laboratory (ORNL).\cite{SEQUOIA1,SEQUOIA2} Measurements were carried out using incident neutron energies of either 30 meV or 120 meV. High-resolution Fermi chopper 2 rotating at frequencies of 360Hz and 600Hz was used for E$_i$=30meV and E$_i$=120meV, respectively. Divergence of incident neutron beam at the sample position due to neutron guide is approximately 20' and 10' for $E_i=$30 meV and $E_i=$120 meV. Scattered neutron travels a distance of 5.5m before reaching the detector array. An energy resolution of $\Delta E/E \sim 1.7\%$ was achieved for both incident energies at elastic energy. The sample was loaded into a flat aluminium container and mounted on a closed cycle cryostat capable of reaching temperatures from 5 K to 550 K.  A series of ``empty can'' data sets were collected under the same experimental conditions and used in background subtraction.

Neutron scattering measurements were also performed using the C5 spectrometer at NRU reactor at Chalk River Laboratories on the same sample.  These measurements were carried out with E$_i$=13.7meV using pyrolytic graphite (PG) as both monochromator and analyzer, with a crystal mosaic of $\sim$33' and $\sim$ 30' respectively.  Natural horizontal and vertical collimations of beam from the reactor is $\sim$33' and $\sim$72'. By using a collimation of [none - 48' - 51' - 144'] along the beam path, an energy and momentum resolution of $\sim$1.2meV and $\sim$0.04$\AA^{-1}$ was achieved at the elastic line.  A PG filter was used to suppress the contamination due to higher order neutrons in the scattered beam.

\begin{figure}[!tb]
	\centering
	\includegraphics[width=0.5\textwidth]{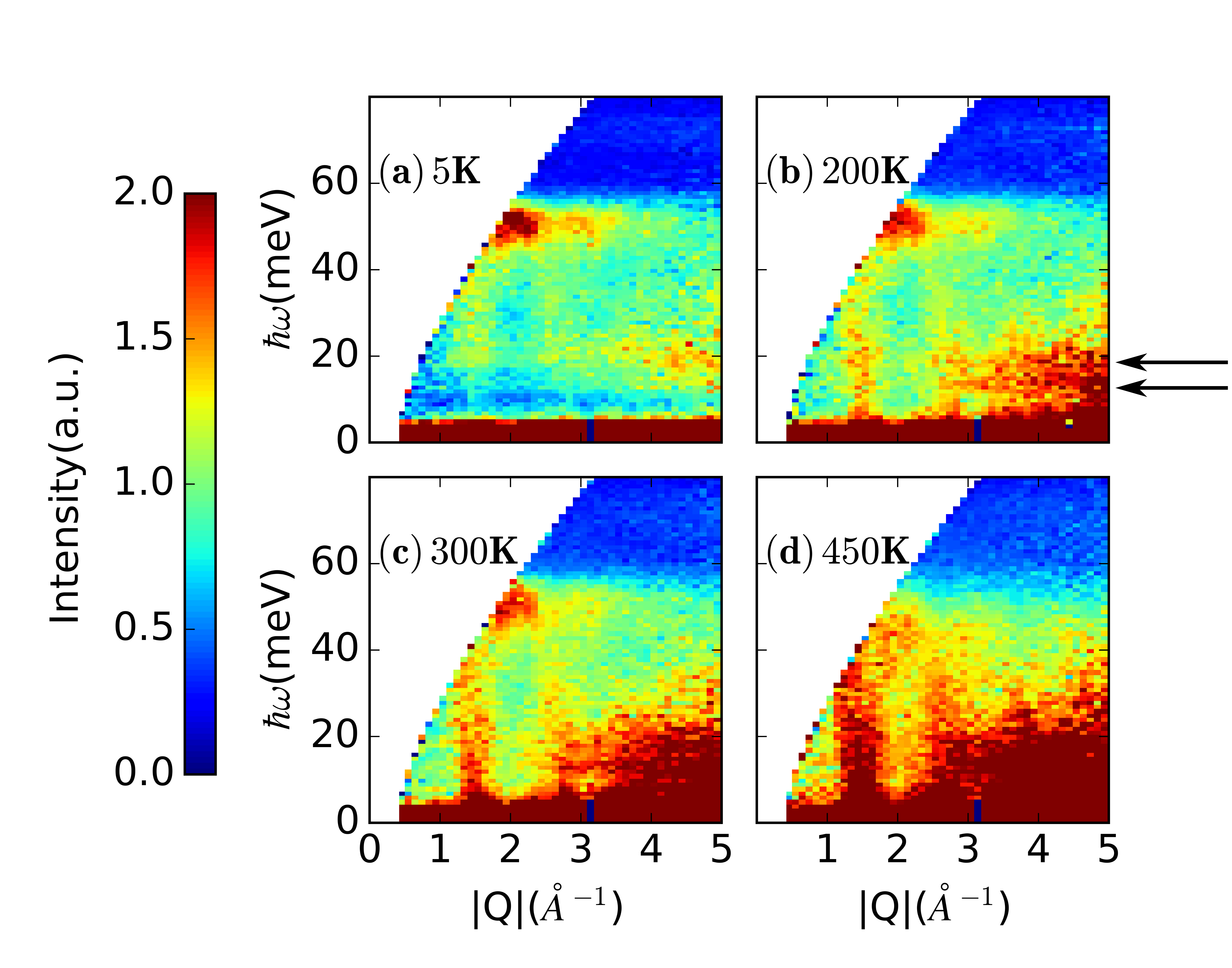}
	\caption{Powder averaged neutron intensity plots measured with incident energy, E$_i$=120meV at various temperatures (a) 5K, (b) 200K, (c) 300K, (d) 450K. The horizontal and vertical axes denote momentum $|\mathbf{Q}|(\AA^{-1})$ and energy transfers $\hbar\omega$(meV). Two optical phonon modes discussed in the main text are indicated by black arrows in (b). Aluminium sample container background has been subtracted from each scan and an arbitrary intensity scale has been used where red (blue) denotes larger (smaller) intensity.}
	\label{120meV}	
\end{figure}

\begin{figure}[htbp]
	\centering
	\includegraphics[width=0.5\textwidth]{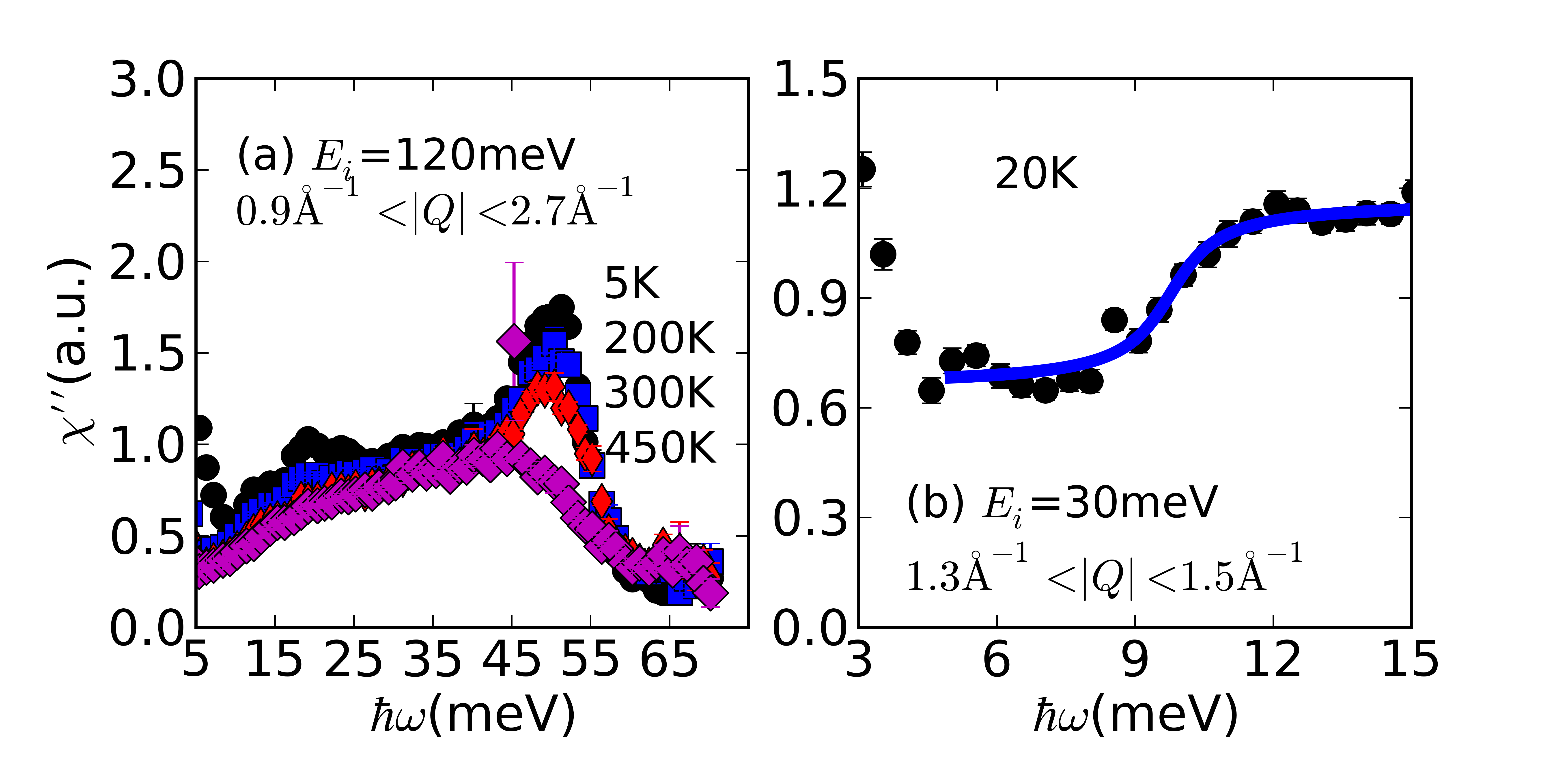}
	\caption{(a) Momentum integrated local susceptibility, $\chi^{\prime\prime}$, at 5K, 200K, 300K, 450K with incident energy E$_i$=120meV and (b) at 20K obtained with incident energy E$_i$=30meV (Fig.~\ref{30meV}(a)). The solid line is a fit to the phenomenological form described in the text. Al sample container background has been subtracted and the same arbitrary intensity scales in Fig.~\ref{120meV} and Fig.~\ref{30meV} have been used for (a) and (b).}
	\label{Qcut}	
\end{figure}

\begin{figure}[!tb]
	\centering
	\includegraphics[width=0.5\textwidth]{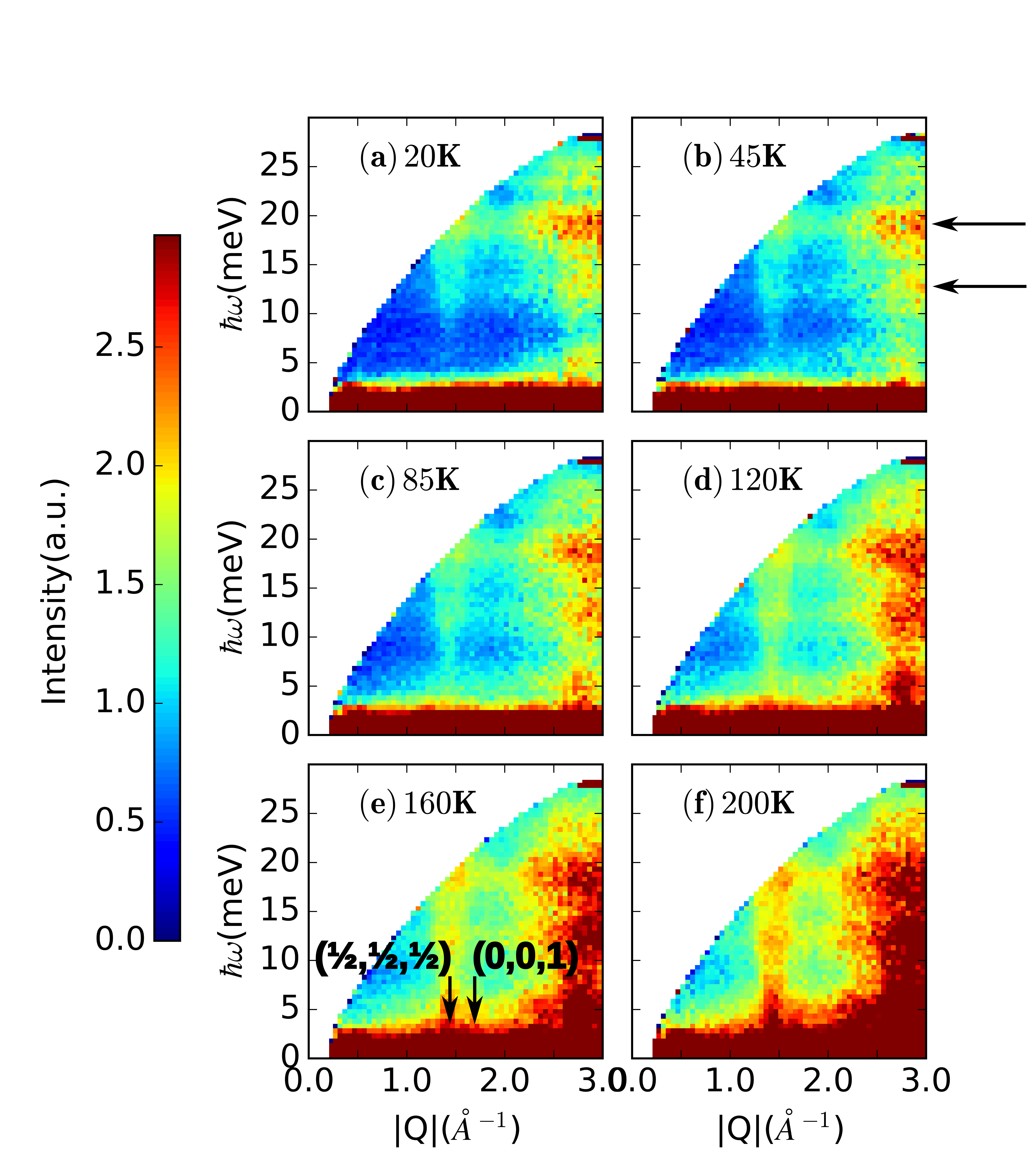}
	\caption{Powder averaged neutron intensity plots with incident energy, Ei=30meV at (a) 20K, (b) 45K, (c) 85K, (d) 120K, (e) 160K, (f) 200K. This data is similar to that of Fig.\ref{120meV}, but obtained with a higher resolution setup. Two modes at $\sim$10meV and $\sim$20meV indicated by black arrows in (b) correspond to the optical phonon modes seen in high E$_i$ data in Fig.\ref{120meV}. Arrows in (e) denote positions for $\mathbf{Q}=(\frac{1}{2},\frac{1}{2},\frac{1}{2})$ and $\mathbf{Q}=(0,0,1)$ in pseudo-cubic notation. Al sample container background has been subtracted from all plots.}
	\label{30meV}	
\end{figure}

\section{Results}

Scattered neutron intensity is plotted as a function of energy transfer, $\hbar\omega=E_i-E_f$ and momentum transfer $|\mathbf{Q}|=|\mathbf{k_i}-\mathbf{k_f}|$ in Fig.~\ref{120meV}(a)-(d) for various temperatures. Highly dispersive magnetic excitations extending up to $\sim$50meV can be clearly resolved in the low $|\bf{Q}|$ region below 3$\AA^{-1}$. Similar dispersive magnetic excitations were observed in Ba$_2$FeReO$_6$, and attributed to powder averaged spin wave modes within linear spin wave theory.\citep{Plumb_BFRO} In addition to magnons, we also observed two modes at $\sim$10meV and $\sim$20meV which are indicated by black arrows in Fig.~\ref{120meV}. They can be attributed to optical phonon modes as their intensities clearly increase with $|\bf{Q}|$. Here we only focus on magnetic excitations in the small $|\bf{Q}|$ region. To compare scattering at different temperatures, inelastic neutron intensity is divided by the Bose-factor defined as $(n(\omega,T)+1)$ to obtain the local susceptibility $\chi^{\prime\prime}(|\mathbf{Q}|,\omega)$. Here $n(\omega,T)=\frac{1}{exp(\hbar\omega/k_B T)-1}$. A large peak of magnon intensity at $\hbar\omega\approx$50meV is revealed by integrating $\chi^{\prime\prime}(|\mathbf{Q}|,\omega)$ within $0.9\AA^{-1}<|\mathbf{Q}|<2.7\AA^{-1}$ as in Fig.~\ref{Qcut}(a) where magnetic contribution dominates. This corresponds to a peak in magnon density of states at magnetic zone boundary energy that results from a Van Hove singularity at the top of magnon band. This peak remains more or less temperature independent up to 300K. A strong damping of the spin wave mode at $\hbar\omega\approx$50meV is observed at 450K, which is clearly shown by a drastic drop in the peak intensity and slight softening in Fig~\ref{Qcut}(a). Since magnon modes are thermally populated when their energies are comparable to thermal energy, zone boundary magnons ($\hbar\omega\sim$50meV) are expected to be strongly damped by magnon-magnon interactions at high temperature ($k_B T\sim$ 50meV or $T\sim$ 500K), in agreement with our observation.

\begin{figure}[!tb]
	\centering
	\includegraphics[width=0.475\textwidth]{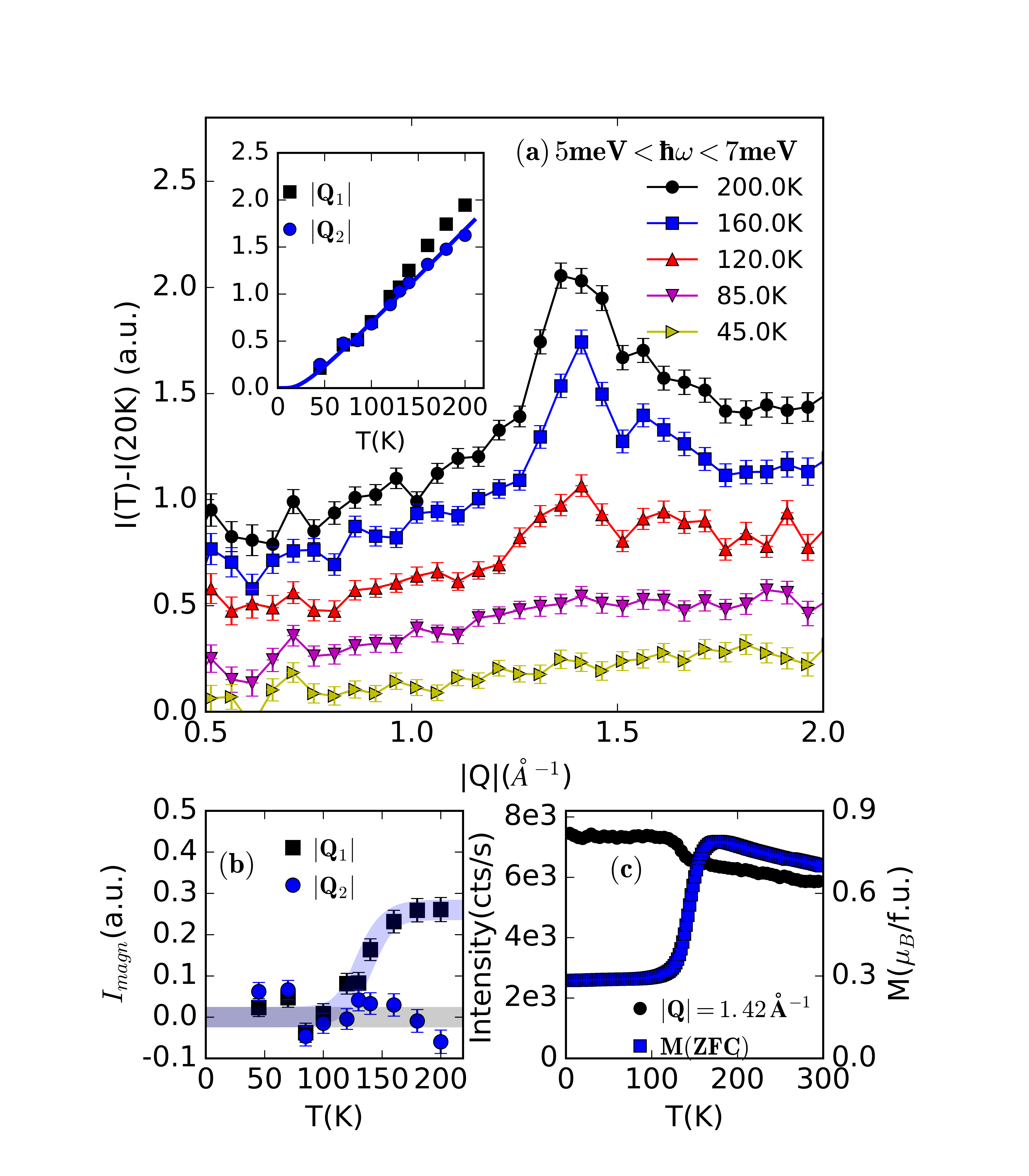}
	\caption{(a) Constant energy cuts below the gap obtained by integrating inelastic neutron intensity in Fig.~\ref{30meV} over energy transfers 5meV$<\hbar\omega<$7meV. 20K data have been substracted from all plots. $\it{Inset}$:  Intensity at $|\mathbf{Q}_1|$ and $|\mathbf{Q}_2|$ as a function of temperature, obtained by integrating constant energy cuts in the main plot within $1.3\AA^{-1}<|\mathbf{Q}|<1.5\AA^{-1}$ and $1.5\AA^{-1}<|\mathbf{Q}|<1.7\AA^{-1}$. Blue solid line shows $n(T)$ for $\hbar\omega$=6meV. It has been scaled to match the data at $|\mathbf{Q}_2|$. (b) Temperature dependence of magnetic inelastic intensity ($I_{magn}$) at $|\mathbf{Q}_1|$ (black square) and $|\mathbf{Q}_2|$ (blue circle). $I_{magn}$ is obtained by subtracting the phonon intensity given by $n(T)$ from integrated intensities in the inset of panel (a). (c) Temperature dependence of the magnetic Bragg peak intensity at $|\bf{Q}_1|$ and Zero-field cooled (ZFC) magnetization of a Ca$_2$FeReO$_6$ pellet. Magnetization was obtained in the presence of an applied field of 0.2T after cooling down to 2K in zero field. Magnetic Bragg peak intensity was obtained at C5 triple axis spectrometer at NRU at fixed $|\bf{Q}|$=1.42$\AA^{-1}$ and $\hbar\omega$=0meV on the same powder sample used for time of flight measurements.}
	\label{Ecut1}
\end{figure}

Interestingly, compared to spin excitation spectrum at 200K, magnon intensity seems to be strongly suppressed for energy transfer $\hbar\omega\lesssim$10meV at 20K. To study the temperature dependence of low energy magnetic excitation in details, we carried out high resolution inelastic neutron scattering experiments with incident energy E$_i$=30meV. Fig.~\ref{30meV} shows representative neutron scattering intensity plots between 20K and 200K, obtained with E$_i$=30meV. With better momentum resolution, one can clearly see that spin wave intensity is strong near $|\mathbf{Q}_1|\sim$1.4$\AA^{-1}$, corresponding to the ordering wave vector at $(\frac{1}{2},\frac{1}{2},\frac{1}{2})$ in pseudo-cubic notation with $a\approx3.8\AA$ (We used pseudo-cubic notation because monoclinic distortion from an ideal cubic unit cell in Ca$_2$FeReO$_6$ is too small to be seen with our experimental momentum resolution). On the other hand, $|\mathbf{Q}_2|\sim$1.6$\AA^{-1}$ corresponds to another Brillouin Zone center (0,0,1) where we observed negligible magnon intensity here. Although both magnetic and structural peaks are allowed at each wave-vector, giving rise to magnon and phonon contributions to the neutron intensity, predominant magnetic contribution near $\mathbf{Q}_1$ is consistent with a stronger magnetic Bragg peak at $\mathbf{Q}_1$. Due to a ferrimagnetic arrangement of Fe and Re moments, denoted by $\mu_{Fe}$ and $\mu_{Re}$ respectively, ratio between the magnetic Bragg peak intensity at $\mathbf{Q}_1$ and $\mathbf{Q}_2$ is proportional to $\left(\frac{\mu_{Fe}+\mu_{Re}}{\mu_{Fe}-\mu_{Re}}\right)^2$, which indicates negligible magnetic contribution to the Bragg intensity at $|\mathbf{Q}_2|$ compared to $|\mathbf{Q}_1|$.\cite{Plumb_BFRO} A better energy resolution also allows us to observe a clear suppression of magnon intensity below $\sim$10meV at T=20K, which shows opening of a spin gap at low temperatures. On the other hand, gapless magnetic excitations are seen in Fig.~\ref{30meV}(e) (160K) and (f) (200K), suggesting closure of the spin gap at high temperatures.

To track detailed temperature dependence of the spin gap, we studied temperature dependence of the magnetic inelastic intensity by making constant energy cuts around 6 meV (integrating over an energy transfer 5meV$<\hbar\omega<$7meV) at various temperatures from T=20K to T=200K. This energy is well below the spin gap (see below), but still higher than the energy resolution. Neutron intensity at a given temperature T and energy transfer $\hbar\omega$=6meV is given by
\begin{equation}
I(T)=I_0+I_{ph}+I_{magn}
\label{I}
\end{equation}
In Eq.~\ref{I}, $I_0$ denotes temperature independent background including elastic tail of incoherent scattering as well as instrumental background. $I_{ph}$ describes phonon scattering in both sample and sample container. Its temperature dependence is well described by Bose factor as $I_{ph}=(n(T)+1)\chi_0$, where $\chi_0$ is the temperature independent local susceptibility due to phonon. Here we have dropped the energy dependence in $n(\omega,T)$ as $\hbar\omega$ was fixed at 6meV for all temperatures. Lastly, $I_{magn}$ is the magnetic inelastic intensity from the sample. To properly account for $I_0$, we subtracted data at 20K from all other temperatures. Resulting difference spectra $\Delta I(T) \equiv I(T)-I(20K)$ are shown in Fig.~\ref{Ecut1}(a). Since $n(\omega,T)\approx0$ at T=20K for $\hbar\omega$=6meV, Eq.~\ref{I} implies $\Delta I(T)=n(T)\chi_0+I_{magn}$. As shown in Fig.~\ref{Ecut1}(a), $\Delta I$ increases gradually with temperature consistent with phonon behaviour for most $|\mathbf{Q}|$'s (See also Supplemental Material for quantitative analysis of the phonon background\cite{SM}). On the other hand, although intensity near the strong magnetic Bragg peak $|\mathbf{Q}_1|$ follows that of phonon background at 45K and 85K, it increases much more rapidly at higher temperatures. Integrated intensity at $|\mathbf{Q}_1|$ is shown in the inset of Fig.~\ref{Ecut1}(a). We also showed intensity at $|\mathbf{Q}_2|$ for comparison. Clearly, $\Delta I$ at $|\mathbf{Q}_2|$ follows the $n(T)$ dependence of phonon background (blue solid line) for all temperatures. In contrast, $\Delta I$ at $|\mathbf{Q}_1|$ rises above the blue line for T$\gtrsim$100K suggesting an increase in neutron intensity due to $I_{magn}$. To obtain $I_{magn}$ at both $|\mathbf{Q}|$'s, phonon background proportional to $n(T)$ was subtracted from $\Delta I$ at each $|\mathbf{Q}|$ and shown in Fig.~\ref{Ecut1}(b). $I_{magn}$ near $|\mathbf{Q}_2|$ remains almost zero for all temperatures. On the other hand, $I_{magn}$ near $|\mathbf{Q}_1|$ is zero for T$<$100K, it increases rapidly going from 100K to 150K and saturates for T$>$150K. As discussed previously, intensity near $|\mathbf{Q}_1|$ ($|\mathbf{Q}_2|$) is dominated by magnon (phonon) due to stronger (weaker) magnetic Bragg peak. Different temperature dependence of $I_{magn}$ near these two $|\mathbf{Q}|$'s unambiguously shows the filling of magnon intensity due to closing of spin gap.

To determine the gap size, neutron intensity in Fig.~\ref{30meV}(a) corrected by Bose factor is integrated from $1.3\AA^{-1}$ to $1.5\AA^{-1}$ around $|\mathbf{Q}_1|$ and plotted as a function of energy transfer as in Fig.~\ref{Qcut}(b). Resulting local susceptibility, $\chi^{\prime\prime}$ is suppressed below $\hbar\omega\sim$10meV indicating the presence of a spin gap.  Data from 5meV to 15meV was fit to a step function $A\tan^{-1}(\frac{\omega-\Delta}{\Gamma})+B$, where $A$, $\Delta$, $\Gamma$ denote the height, center and width of the step function and $B$ denotes the background. The gap size, defined as center of the step function $\Delta$ is determined to be 9.8(5)meV.

Drastic change in low energy magnetic dynamics is also reflected in the temperature dependence of magnetic Bragg peak intensity, which is proportional to the square of static moment size. As shown in Fig.~\ref{Ecut1}(c), the magnetic Bragg peak intensity at $|\mathbf{Q}_1|$ remains constant for T$<$100K. It drops rapidly as temperature increases from 100K to 150K and continues to decrease gradually for T$>$150K. Gradual decrease in magnetic Bragg intensity for T$>$150K is consistent with the observed spin gap closing, since increases in spin fluctuation with thermal population of low energy magnons would reduce the static moments with increasing temperature. On the other hand, magnetic Bragg intensity remains unchanged for T$<$100K since low energy spin fluctuations are pre-empted by the presence of a large spin gap. The abrupt $\sim15\%$ drop in $|\mathbf{Q}_1|$ Bragg peak intensity from 100K to 150K shown in Fig.~\ref{Ecut1}(c) was also observed in the neutron diffraction study by Oikawa $\it{et.\,al}$\cite{CFROstructuraldistortion} and was attributed to re-orientation of magnetic moments due to orbital ordering from Rietveld analysis.

Since isotropic Heisenberg Hamiltonian is rotationally invariant, magnon spectrum should be gapless by Goldstone theorem. Presence of magnon gap at low temperature thus requires single ion anisotropy or anisotropic exchange terms that break full rotational symmetry of the magnetic Hamiltonian. Gapless spectra observed in the high temperature phase of Ca$_2$FeReO$_6$ as well as in Ba$_2$FeReO$_6$ suggest removal of the anisotropy terms. This picture is consistent with bulk magnetization data shown in Fig.~\ref{Ecut1}(c). For T$<$100K, ordered moments are confined to crystallographic directions favoured by the anisotropy terms, which are randomly oriented in a powder sample and lead to small overall magnetization when field is not too large. For T$>$150K, the absence of anisotropy means that ordered moments can be freely aligned by an external field, leading to the observed large increase in magnetization. A significant drop in coercive field was also observed going from 100K to 150K, consistent with a decrease in magnetic anisotropy. \citep{CFRO_MvH}

\section{Discussions}
Our neutron scattering results show that Ca$_2$FeReO$_6$ exhibits a well-defined spin wave excitation in the magnetically ordered phase. The spin excitation is gapless for T$\gtrsim$150K and becomes gapped (with gap size$\sim$10meV) below $\sim$150K coinciding with the reported metal insulator transition\citep{CFRO_MIT} and onset of structural distortion\citep{CFROstructuraldistortion}. The structural distortion has been interpreted as a consequence of Re orbital ordering both by considering the effects of ReO$_6$ octahedra distortion on local crystal field levels \citep{CFROstructuraldistortion} as well as by a DFT calculation \citep{CFRO_DFT3}.

Since distortion of the orbital electron density is more pronounced in the presence of strong orbital ordering, one expects larger low symmetry magnetic interactions in the orbitally ordered phase due to spin orbit coupling (SOC). This is entirely consistent with the gap opening shown by our neutron scattering results at low temperatures. Therefore, our results provide compelling evidence for orbital order in Ca$_2$FeReO$_6$ for T$\lesssim$150K. Quantitatively, inclusion of SOC in DFT calculations in the orbitally ordered phase give rise to easy axis anisotropy along $b$ with a magnetic anisotropy energy per Re atom of $\sim$6meV \citep{CFRO_DFT3}, in rough agreement with the gap size ($\sim$10meV) observed in our data. Slight structural distortion was also observed in Ba$_2$FeReO$_6$ below the magnetic transition temperature\citep{BFRO_OO}. Unlike Ca$_2$FeReO$_6$, its magnon spectrum remains gapless throughout the magnetically ordered phase\cite{Plumb_BFRO}. Comparison between Ca$_2$FeReO$_6$ and Ba$_2$FeReO$_6$ therefore indicates that no actual orbital order occurs in Ba$_2$FeReO$_6$. Stronger tendency of orbital order in Ca$_2$FeReO$_6$ was attributed to a larger tilt of ReO$_6$ octahedra by Lee and Marianetti\cite{CFRO_DFT3}.

Separate magnetic and orbital ordering transitions in Ca$_2$FeReO$_6$ as well as magnetic order without orbital order in Ba$_2$FeReO$_6$ indicate a weak coupling between spin and orbital degrees of freedom in A$_2$FeReO$_6$. This is somewhat surprising given the large SOC on Re ions. In a recently popular atomic picture where local physics is determined by SOC and crystal field, Re spin and orbital angular momenta are locked into $J_{eff}$=2 local moments that interact through highly anisotropic interactions as shown theoretically by Chen and Balents\citep{d2theory}.

Although experimental signatures for anisotropic interactions are largely lacking in rhenates, gapped magnon spectra consistent with anisotropic interactions between $J_{eff}$ moments have been observed in many osmate and iridate compounds with chain\cite{Sr3CuIrO6, Sr3NiIrO6_INS, Sr3NiIrO6_RIXS}, layered\citep{Sr327,Sr214_inplane}, perovskite\citep{NaOsO3} and double perovskite structures \citep{IrDPINS_gap, Ba2YOsO6, Sr2ScOsO6, Sr2MgOsO6, marjerrisoncasey2016}. Among the double perovskites studied, Sr$_2$ScOsO$_6$ (with magnon gap $\Delta$=12meV)\cite{Sr2ScOsO6}, Sr$_2$MgOsO$_6$ ($\Delta$=7meV)\cite{Sr2MgOsO6}, Ba$_2$ZnOsO$_6$ ($\Delta$=7meV) and Ba$_2$MgOsO$_6$ ($\Delta$=12meV) \cite{marjerrisoncasey2016} contain Os$^{6+}$ ion with the same electronic configuration as Re$^{5+}$ and hence directly indicate the importance of anisotropic interactions between $5d^{2}$ ions.

Gapless magnetic excitations observed in magnetically ordered phases therefore suggest a sub-dominant role of SOC in A$_2$FeReO$_6$. This also explains the separation of spin and orbital degrees of freedom in A$_2$FeReO$_6$. Theoretically, Lee and Marianetti\cite{CFRO_DFT3} showed that electronic properties of A$_2$FeReO$_6$ were determined mainly by electronic correlation and lattice distortion and were not qualitatively changed by SOC. This is consistent with the discussions based on our experimental observation. Small effects of SOC in Ba$_2$FeReO$_6$ have been inferred from a branching ratio of $\sim$2 determined by X-ray absorption spectroscopy\cite{BR_iridate}, which is similar to the statistical value expected in the absence of SOC\cite{BR_theory}. Qualitatively, small effects of SOC in A$_2$FeReO$_6$ might be explained by a large Re bandwidth due to strong hybridization with surrounding Fe ions. A large electronic bandwidth mixes different $J_{eff}$ levels and weaken the effects of SOC.  Breakdown of $J_{eff}$ picture is also supported by resonant inelastic X-ray scattering measurement on A$_2$FeReO$_6$ reporting highly damped intra $t_{2g}$ excitations.\citep{ReDP_RIXS}

We note that a recent inelastic neutron scattering on another $3d-5d$ DP with magnetic $3d$ ion, Sr$_2$FeOsO$_6$\citep{Sr2FeOsO6_INS} also observed a gap opening below a secondary phase transition, somewhat similar to Ca$_2$FeReO$_6$. However, gap opening in Sr$_2$FeOsO$_6$ is accompanied by a magnetic structure change\citep{Sr2FeOsO6_structure}, which is different from Ca$_2$FeReO$_6$ where magnetic order remains unchanged. In addition, OsO$_6$ octahedra actually become more symmetrical in the low temperature phase of Sr$_2$FeOsO$_6$.\citep{SFOO_localstructure} Therefore, the origin of the gap opening in Sr$_2$FeOsO$_6$ is different from that in Ca$_2$FeReO$_6$.

So far we have focussed on the low energy part of magnetic excitations in Ca$_2$FeReO$_6$. We now move on to discuss its high energy part. One puzzling feature of high energy spin excitations is the observation of only one strong magnon band at $\hbar\omega\approx$50meV, as shown in Fig.~\ref{120meV}(a). Since Ca$_2$FeReO$_6$ consists of Fe and Re moments of different sizes, one expects two magnon modes and hence two strong bands of inelastic intensity coming from zone boundary magnons within ferrimagnetic spin wave theory. One possibility is an accidental merging of Fe and Re magnon bands at the zone boundary. In the Appendix, we showed that this is unlikely because we did not observe splitting of the two bands by the anisotropy terms responsible for opening a spin gap at low temperatures. Another possibility is that the observed spin wave only comes from Fe local moments while dynamics of Re electrons are much faster and can be integrated out. This $3d$-spin only model has been proposed for Sr$_2$FeMoO$_6$ and Sr$_2$CrOsO$_6$\cite{SFMO_theory,SCOO_theory}. Since Fe magnetic moments are ferromagnetically ordered, we expect the spin wave intensity to be strong at both `antiferromagnetic'-$\bf{Q}$ $(\frac{1}{2},\frac{1}{2},\frac{1}{2})$ and `ferromagnetic'-$\bf{Q}$ $(0,0,1)$ in a Fe-spin only model. However, we only observed strong magnetic inelastic intensity at $(\frac{1}{2},\frac{1}{2},\frac{1}{2})$ as shown in Fig.\ref{30meV}. Therefore, we conclude that Re magnetic moments must contribute to spin dynamics in Ca$_2$FeReO$_6$. The most plausible explanation for observation of only one magnon band is therefore selective damping of Re magnon. This might be due to strong charge fluctuation on the Re sites, which is expected from the close proximity of Re electrons to a Mott instability. On the other hand, localized Fe moments give rise to a well-defined spin wave mode seen at $\sim$50meV.

\section{Conclusions}
In summary, we have carried out inelastic neutron scattering studies on Ca$_2$FeReO$_6$. We found well-defined spin wave excitations that are gapless for T$\gtrsim$150K but become gapped at $\sim$150K. This provides strong dynamical evidence for orbital ordering below $\sim$150K in Ca$_2$FeReO$_6$ proposed by a previous structural study\cite{CFROstructuraldistortion} and recent DFT+U calculations\citep{CFRO_DFT3}. On the other hand, no spin gap was found in Ba$_2$FeReO$_6$\citep{Plumb_BFRO} which indicates the absence of such orbital ordering, consistent with theoretical predictions. We argued that unlike other $5d$ materials where spin-orbit coupling is essential in describing the magnetic dynamics, SOC plays a sub-dominant role in describing the magnetism of A$_2$FeReO$_6$.

Moreover, we found only one zone boundary magnon band in Ca$_2$FeReO$_6$ which strongly contradicts the simple ferrimagnetic spin wave picture. This points to a co-existence of undamped Fe and strongly damped Re magnon modes. To test this possibility, future polarized neutron or resonant inelastic X-ray scattering on single crystals of Ca$_2$FeReO$_6$ that can resolve Fe and Re magnon modes are desirable.

\begin{acknowledgements}
Work at the University of Toronto was supported by the Natural Science and Engineering Research Council (NSERC) of Canada. B. C. J. and T. W. N. are supported by the Research Center Program of IBS (Institute for Basic Science) in Korea (IBS-R009-D1).
B.Y. would like to acknowledge support from the Ontario Graduate Scholarship. C. W. and N. H. are supported by Basic Science Research Program through the National Research Foundation of Korea(NRF) funded by the Ministry of Education, Science and Technology (2013R1A1A2009777). Use of the Canadian Neutron Beam Centre at Chalk River Laboratories is supported by the National Research Council (NRC) and Atomic Energy of Canada Limited (AECL).  Use of the Spallation Neutron Source at Oak Ridge National Laboratory is supported by the Scientific User Facilities Division, Office of Basic Energy Sciences, U.S. Department of Energy.
\end{acknowledgements}
\section{Appendix}

\subsection{Spin wave analysis}

\begin{figure}[!tb]
	\centering
	\begin{subfigure}
	\centering
		\includegraphics[width=0.5\textwidth]{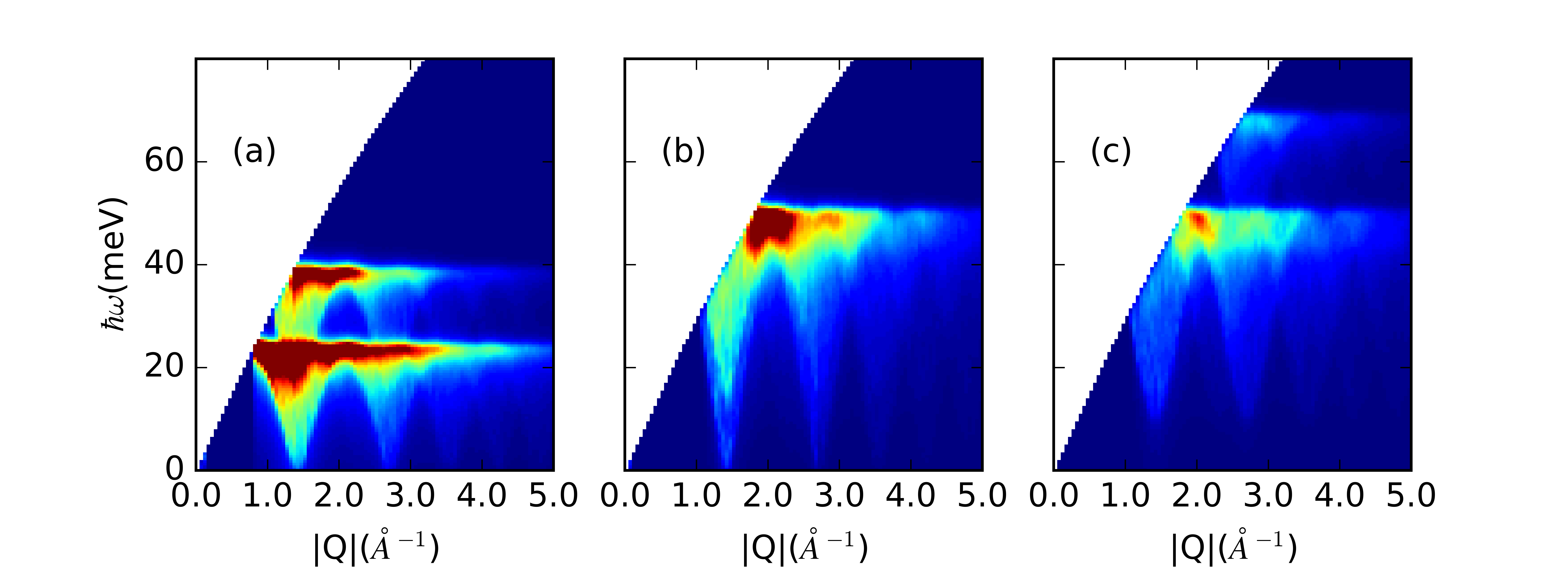}
	\end{subfigure}
	\begin{subfigure}
	\centering
		\includegraphics[width=0.5\textwidth]{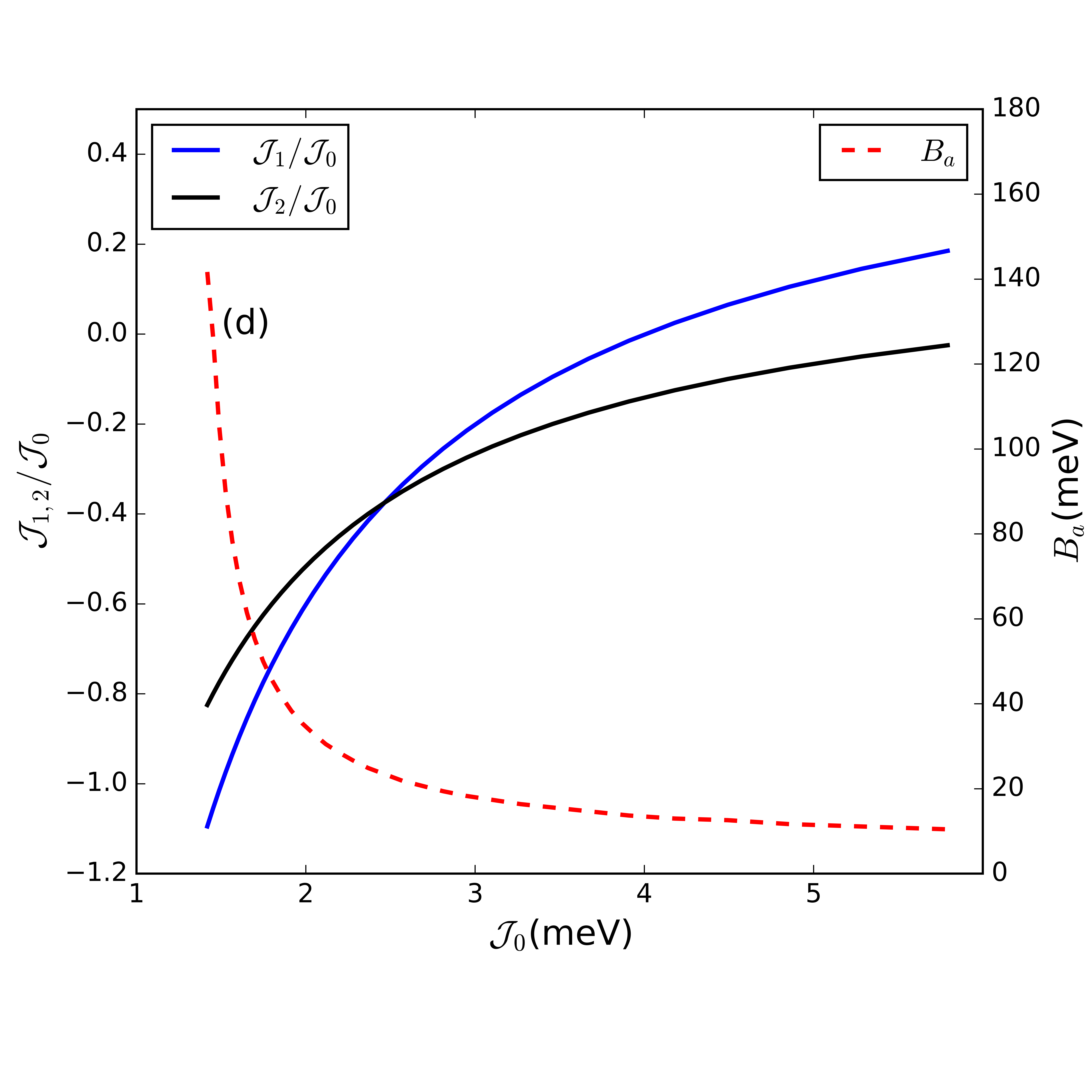}
	\end{subfigure}
	\caption{(a)-(c): Simulated powder averaged spin wave spectra using $\mathcal{F}$=2.1 and $\mathcal{R}$=1.3 for (a) $\mathcal{J}_0$=3.1meV, $\mathcal{J}_1=\mathcal{J}_2=B_a$=0, (b) $\mathcal{J}_0$=2.9meV, $\mathcal{J}_1$=-0.7meV, $\mathcal{J}_2$=-0.8meV, $B_a$=0 and (c) $\mathcal{J}_0$=2.9meV, $\mathcal{J}_1$=-0.7meV, $\mathcal{J}_2$=-0.8meV, $B_a$=18.6meV. Calculation of powder averaged spectra is done using SpinW package.\citep{SpinW} All spectra have been corrected for magnetic form factors and instrumental resolutions. (d): $\it{left\,axis}$:$\{\mathcal{J}_0,\mathcal{J}_1,\mathcal{J}_2\}$ parameters constrained by setting zone boundary energies of both modes in the local moment model to be 50meV. Horizontal axis denotes the absolute magnitude of nearest Fe-Re interaction, $\mathcal{J}_0$. Vertical axis denotes the magnitude of next nearest neighbour Re-Re and Fe-Fe interaction relative to the nearest neighbour interaction, $\mathcal{J}_1/ \mathcal{J}_0$ and $\mathcal{J}_2/ \mathcal{J}_0$. $\it{right\,axis}$: The effective anisotropy field $B_a$ required to produce a gap of 10meV for each set of $\{\mathcal{J}_0,\mathcal{J}_1,\mathcal{J}_2\}$}
	\label{calculation1}
\end{figure}

In this section, we explore the potential merging of Fe and Re magnon modes in a local moment model. The Fe/Re local moment model used here is similar to that considered by Plumb $\it{et.\,al}$\cite{Plumb_BFRO}. In addition to the nearest neighbour Fe-Re coupling $\mathcal{J}_0$, we also included next nearest neighbour Re-Re and Fe-Fe couplings, $\mathcal{J}_1$ and $\mathcal{J}_2$
\begin{equation}
\begin{split}
\mathcal{H}=\mathcal{J}_0 \sum_{\langle i,j\rangle}\vec{\mathcal{F}_i}\cdot\vec{\mathcal{R}_j}+\mathcal{J}_1 \sum_{\langle i, i^{\prime}\rangle}\vec{\mathcal{R}}_i\cdot\vec{\mathcal{R}}_{i^\prime}\\
+\mathcal{J}_2 \sum_{\langle j, j^{\prime}\rangle}\vec{\mathcal{F}}_j\cdot\vec{\mathcal{F}}_{j^\prime}-B_a\sum_{i} \mathcal{R}_{z,i}
\end{split}
\end{equation}
where $\vec{\mathcal{R}}_i$ and $\vec{\mathcal{F}}_j$
denote effective Re- and Fe local moments respectively. In A$_2$FeReO$_6$, long range Re-Re and Fe-Fe magnetic interactions are expected to be significant due to large electronic itinerancy at the proximity of a metal insulator transition. An anisotropy term of the form $B_a \mathcal{R}_z$ has been added to account for the spin gap observed at low temperatures. It is equivalent to Ising anisotropy of the form $\mathcal{J}_z \mathcal{R}_{z,i} \mathcal{R}_{z,i^\prime}$ and single ion anisotropy $D \mathcal{R}_z^2$ within linear spin wave theory. Since terms like these come from strong spin orbit coupling on Re, it is not included for the Fe moment, which is orbitally inactive with $S=\frac{5}{2}$. Energies of the two magnon modes are\cite{Lovesey} 
\begin{equation*}
2\hbar \omega_{\pm} (\mathbf{Q})=2\Omega ( \mathbf{Q}) \pm (B_a+b_1( \mathbf{Q} )-b_2( \mathbf{Q} ))
\end{equation*}

, where we defined the following terms: 

\begin{equation*}
b_1( \mathbf{Q} )= \mathcal{F} \mathcal{J}_0(\mathbf{0})-\mathcal{R} (\mathcal{J}_1( \mathbf{0})-\mathcal{J}_1( \mathbf{Q} ))
\end{equation*}

\begin{equation*}
b_2(\mathbf{Q})= \mathcal{R} \mathcal{J}_0(\mathbf{0})-\mathcal{F}( \mathcal{J}_2( \mathbf{0} )-\mathcal{J}_2( \mathbf{Q}))
\end{equation*}

\begin{equation*}
2\Omega ( \mathbf{Q} )=[(B_a+b_1( \mathbf{Q} )+b_2( \mathbf{Q} ))^2-\mathcal{R} \mathcal{F}(2 \mathcal{J}_0( \mathbf{Q} )^2)]^{ \frac{1}{2} }
\end{equation*} 
Here we define $\mathcal{J}_{\alpha=0,1,2}(\mathbf{Q})=\mathcal{J}_\alpha\sum_\mathbf{\delta} exp(i\mathbf{\delta}\cdot \mathbf{Q})$, where $\mathbf{\delta}$ denotes all neighbouring sites coupled to the $i^{th}$ site by interaction $\mathcal{J}_\alpha$.

First we set $B_a=0$, which corresponds to the high temperature phase of Ca$_2$FeReO$_6$, two well-separated magnon modes are predicted by linear spin-wave theory with only nearest neighbour interactions. This was demonstrated by Plumb $\it{et.\,al}$\cite{Plumb_BFRO}. Calculated spin wave spectrum from ref[\onlinecite{Plumb_BFRO}] using $\mathcal{F}=2.1$, $\mathcal{R}=1.3$ and $\mathcal{J}_0$=3.1meV is reproduced in Fig.~\ref{calculation1}(a).

The lower magnon branch can be moved up and made to coincide with the upper branch by turning on next nearest neighbour interactions. The observed magnon spectrum in Ca$_2$FeReO$_6$ can be captured qualitatively by setting zone boundary energies of both branches to be 50meV, which generates a range of allowed $\{\mathcal{J}_0,\mathcal{J}_1,\mathcal{J}_2 \}$ depending only on the value of $\mathcal{J}_0$ as shown in Fig.~\ref{calculation1}(d). The calculated spectrum using one set of the allowed parameters $\mathcal{J}_0$=2.9meV, $\mathcal{J}_1$=-0.7meV, $\mathcal{J}_2$=-0.8meV is shown in Fig.~\ref{calculation1}(b). For comparison, we used the same values of $\mathcal{F}$ and $\mathcal{R}$ as in Plumb $\it{et\,al.}$\cite{Plumb_BFRO}. 

Turning to the low temperature data, we observed the spin gap to be 10meV which sets a constraint on $B_a$. As shown in Fig.~\ref{calculation1}(d), a large $B_a>$10meV is required to reproduce the observed spin gap for the entire range of interactions considered. Since degeneracy at the zone boundary is accidental, introducing a large anisotropy term splits the two zone boundary modes by an amount $\hbar\omega_+-\hbar\omega_-=B_a$. This is explicitly shown in Fig.~\ref{calculation1}(c) for the interaction parameters used in Fig.~\ref{calculation1}(b). Such splitting is not observed in our data at low temperatures, which points towards the inadequancy of the above spin wave analysis. 

Another way to merge Fe and Re magnon modes in a local moment model is to use the same magnetic moment sizes for both Re and Fe. Similar to the above discussions, we expect degeneracy of the two modes to be broken by anisotropy terms included only for Re sites at low temperatures. Moreover, identical magnetic moment sizes are ruled out by powder neutron diffraction\citep{CFROstructuraldistortion} which refined Fe and Re moments to be $\sim$4$\mu_B$ and $\sim$1$\mu_B$ respectively.

\end{document}